\documentclass{jpp}
\usepackage{subeqn}
\usepackage{epsfig}

\let\realverbatim\verbatim
\let\realendverbatim\endverbatim
\renewenvironment{verbatim}
  {\par\addvspace{6pt plus 1pt minus 2pt}\realverbatim}
  {\realendverbatim\addvspace{6pt plus 1pt minus 2pt}}

\ifprodtf \else
  \checkfont{eurm10}
  \iffontfound
    \IfFileExists{upmath.sty}
      {\typeout{^^JFound AMS Euler Roman fonts on the system,
                   using the 'upmath' package.^^J}%
       \usepackage{upmath}}
      {\typeout{^^JFound AMS Euler Roman fonts on the system, but you
                   don't seem to have the}%
       \typeout{'upmath' package installed. JPP.cls can take advantage
                 of these fonts,^^Jif you use 'upmath' package.^^J}%
       \providecommand\umu{\umu}%
      }
  \else
    \providecommand\umu{\mu}%
  \fi
\fi

\ifprodtf \else
  \checkfont{msam10}
  \iffontfound
    \IfFileExists{amssymb.sty}
      {\typeout{^^JFound AMS Symbol fonts on the system, using the
                'amssymb' package.^^J}%
       \usepackage{amssymb}%

      }{}
  \else
  \fi
\fi

\ifprodtf \else
  \IfFileExists{amsbsy.sty}
    {\typeout{^^JFound the 'amsbsy' package on the system, using it.^^J}%
     \usepackage{amsbsy}}
    {}
\fi

\newcommand{\be}{\begin{equation}}
\newcommand{\ee}{\end{equation}}

\newcommand{\eqa}{\begin{eqnarray*}}
\newcommand{\eqe}{\end{eqnarray*}}
\newcommand{\eqnu}{\begin{eqnarray}}
\newcommand{\eqne}{\end{eqnarray}}

\newcommand{\eeq}{\end{eqnarray}}  

\newdefinition{definition}[theorem]{Definition}

\title[Journal of Plasma Physics]{
Simulations of 2D magnetic electron drift vortex mode turbulence in plasmas}

\author[Dastgeer Shaikh \& PK Shukla]
{D\ls A\ls S\ls T\ls G\ls E\ls 
E\ls R\ns S\ls H\ls A\ls I\ls K\ls H\ls and\ls P\ls K\ls S\ls H\ls U\ls K\ls \L\ls A\ls$^\dagger$}
\affiliation{Institute of Geophysics and Planetary Physics (IGPP),\\
University of California, Riverside, CA 92521. USA.\\
{\tt Email:dastgeer@ucr.edu}\\
$^\dagger$Institut f\"ur Theoretische Physik IV, Fakult\"at f\"ur Physik und
Astronomie, Ruhr-Universit\"at Bochum, D-44780 Bochum, Germany\\
{\tt Email: ps@tp4.rub.de}}
\date{31 March 2008; Revised on 11 April 2008}
\pubyear{2008} 
\volume{00}
\part{0}
\pagerange{\pageref{firstpage}--\pageref{lastpage}}
\doi{S0963548301004989}

\begin{document}

\label{firstpage}
\maketitle

\begin{abstract}
Simulations are performed to investigate turbulent properties of nonlinearly 
interacting two-dimensional (2D) magnetic electron drift vortex (MEDV) modes
in a nonuniform unmagnetized plasma. The relevant nonlinear equations 
governing the dynamics of the MEDV modes are the wave magnetic field 
and electron temperature perturbations in the presence of the equilibrium
density and temperature gradients. The important nonlinearities come from the
advection of the electron fluid velocity perturbation and the electron temperature,
as well as from the nonlinear electron Lorentz force. Computer simulations of 
the governing equations for the nonlinear MEDV modes reveal the generation 
of streamer-like electron flows, such that the corresponding gradients in 
the direction of the inhomogeneities tend to flatten out. By contrast, the 
gradients in an orthogonal direction vary rapidly.  Consequently, the inertial 
range energy spectrum in decaying MEDV mode turbulence exhibits a much steeper 
anisotropic spectral index. The magnetic structures in the MEDV mode turbulence 
produce nonthermal electron transport in our nonuniform plasma.
\end{abstract}


\section{Introduction}

About a quarter century ago, several authors [\cite{r1,r2,r3}]
predicted the existence of the magnetic-electron-drift vortex (MEDV)
mode in a nonuniform unmagnetized plasma because of its possible
importance in laser fusion as well as in nonstationary processes in
laboratory plasmas.  The MEDV mode is purely magnetic and has
negligible density and electrostatic potential perturbations in an
inhomogeneous plasma with fixed ion background. When the equilibrium
electron density and electron temperature gradients are simultaneously
present, the MEDV mode gets unstable [\cite{r2,r3}] because of a first
order baroclinic force, which can enhance the wave magnetic field and
electron temperature perturbations if the electron temperature
gradient is sufficiently large in comparison with the density
gradient. Large amplitude MEDV modes are subjected to parametric
instabilities [\cite{r4,r5}] involving the decay and modulational
interactions [\cite{r6,r7,r8}]. Due to the parametric processes the
magnetic energy of the MEDV modes is redistributed among other modes
in plasmas.

In their classic paper, Nycander {\it et al.} [\cite{r9a}] derived
a pair of equations governing the dynamics of nonlinearly interacting
MEDV modes.  It turns out that the nonlinear mode coupling equations
contain the intrinsic Jacobean nonlinearities arising from the
advection of the electron fluid velocity and electron temperature
perturbations, as well as the nonlinear Lorentz force. These
nonlinearities are responsible for the formation of coherent magnetic
structures in the form of a dipolar vortex [\cite{r9,r10}] and a vortex
street [\cite{r11}].

Recently, there has been a renewed interest in the study of MEDV mode
turbulence [\cite{r11,r12}]. Specifically, Jucker and
Pavlenko [\cite{r11}] and Andrushchenko {\it et al.} [\cite{r12}] have
presented analytical studies of the generation of large-scale magnetic
fields (magnetic streamers) via the modulational instability of
coherent and partially incoherent MEDV modes. The use of a
quasi-linear theory also reveals the shearing of micro-turbulence by
the electron flows and the corresponding electron diffusion in the
presence of zonal magnetic fields and magnetic streamers [\cite{r12}].

In this paper, we present results of computer simulations of 2D
nonlinearly interacting MEDV modes whose dynamics is governed by the
nonlinear equations in Ref. [\cite{r9}].  Needless to say, we
numerically solve the coupled equations for the wave magnetic field
and the electron temperature perturbation in order to investigate the
formation of magnetic and temperature fluctuation structures, as well
as the associated turbulent spectra and the resultant electron
transport in the magnetic structures.

\section{2D nonlinear equations}

We consider a nonuniform plasma in the presence of the equilibrium
density and electron temperature gradients. The latter, which are
along the $x$ axis in a Cartesian coordinate system, are maintained by
the external sources, such as the dc electric field $E_{0x}$. The
dynamics of the MEDV mode in our nonuniform plasma is governed by the
electron momentum equation
$$
\left(\frac{\partial}{\partial t} + {\bf u} \cdot \nabla\right){\bf u}
+ \frac{e}{m}\left({\bf E} + \frac{1}{c}{\bf u} \times {\bf B}\right) +
\frac{1}{mn}\nabla (nT)=0,
\eqno(1)
$$
the electron energy equation
$$
\left(\frac{\partial}{\partial t} + {\bf u} \cdot \nabla\right)T + (\gamma-1) T \nabla \cdot {\bf u} =0,
\eqno(2)
$$
the Faraday law
$$
\nabla \times {\bf E} =-\frac{1}{c}\frac{\partial {\bf B}}{\partial t},
\eqno(3)
$$
and the Amp\`er\'es law
$$\nabla \times {\bf B} = - \frac{4\pi e n {\bf u}}{c},
\eqno(4)
$$
where ${\bf u}$ is the electron fluid velocity, $T$ is the electron
temperature, $e$ is the magnitude of the electron charge, $m$ is the
electron mass, $c$ is the speed of light in vacuum, $\gamma (=2/3)$ is
the adiabatic index (the ratio of the specific heat), and ${\bf E}$
and ${\bf B}$ are the electric and magnetic fields, respectively. The
displacement current in (4) has been neglected since the MEDV mode
phase speed is much smaller than $c$. The assumption of immobile is
justified as long as the MEDV mode frequency is larger than the ion
plasma frequency.

Taking the curl of (1), using (3) and (4) and letting ${\bf B} = \hat
{\bf z} B$, $T = T_0(x) + T_1$, where $\hat {\bf z}$ is the unit
vector along the $z$ axis and $T_1 (\ll T_0)$ is a small temperature
fluctuation in the equilibrium value $T_0 (x)$, we obtain [\cite{r9}]

$$\frac{\partial}{\partial t} \left(B -\lambda^2 \nabla^2 B\right) 
+ \beta \frac{\partial T_1}{\partial y} + \frac{e\lambda^4}{m c}[B,\nabla^2B], 
\eqno(5)
$$
where $\beta = cK_n/e$, $K_n =\partial {\rm ln} n (x)/\partial x$, $\lambda =c/\omega_p$,
and $\omega_p =(4\pi n e^ 2/m)^ {1/2}$ is the electron plasma frequency. We have denoted
$\nabla^ 2 = (\partial^ 2/\partial x^ 2)+ (\partial^ 2/\partial z^ 2)$ and $[F,G] =
\hat {\bf z} \times \nabla F \cdot \nabla F$ represents the Jacobean nonlinearity.

The electron energy equation (2) is written as
$$\frac{\partial T_1}{\partial t} 
+  \alpha \frac{\partial B}{\partial y} + \frac{e\lambda^2}{m c}[B,T] =0,
\eqno(6)
$$
where $\alpha=\lambda^2 (eT_0/mc)[(2/3)K_n -K_T]$, and $K_T =|\partial {\rm ln} T_0/
\partial x|$.

In the linear limit, we neglect the nonlinear terms in (5) and (6), and by assuming that 
$B$ and $T_1$ are proportional to $\exp(-i\omega t + i {\bf k} \cdot {\bf r})$, we Fourier transform
and combine the resultant equations to obtain the MEDV mode frequency 
$$\omega = k_y\sqrt{\alpha \beta/(1+k^2 \lambda^2)}, 
\eqno(7)$$
where $\alpha \beta =V_T^2 \lambda^2 K_n [(2/3) K_n -K_T]$ and
$V_T =(T/m)^{1/2}$ is the electron thermal speed. Furthermore, $k^ 2 = k_y^ 2 + k_x^ 2$
and $k_y (k_x)$ is the component of the wave vector ${\bf k}$ along the $y (x)$ axis.
The wavelength of the MEDV mode is assumed to be much smaller than $K_n^ {-1}$ and $K_T^ {-1}$. 
Equation (7) reveals that for $k^ 2 \lambda^ 2 \ll 1$, the phase speed $\omega/k_y$ of 
the MEDV mode is constant, while for $k^ 2 \lambda^ 2 \gg  1$, we have $\omega =
(k_y/k)\sqrt{\alpha \beta/\lambda^ 2}$. The latter indicates frequency condensation 
at $k_x \gg k_y$.

Equations (5) and (6) admit the energy integral
$$E =\int [B^2 + \lambda^2 (\nabla B)^2 + (\beta/\alpha)T_1^2] dx dy.
\eqno(8)
$$
The total energy above comprises the magnetic field, magnetic field
vorticity and temperature energies. The energy integral (8) is used to
check the numerical validity of our code, which must be preserved under ideal 
conditions.

\section{Simulation results}

We suitably normalize (e.g. $t$ by $\omega_p^{-1}$, $\nabla$ by
$\lambda^{-1}$, $B$ by $mc\omega_p/e$, $T_1$ by $T_0$ ) (5) and (6)
and investigate numerically the properties of 2D localized magnetic
structures, as well as examine the associated magnetic turbulence
spectrum and nonthermal cross-field electron transport caused by the
magnetic structures.

We have developed a code to integrate the normalized (as mentioned
above) equations (5) and (6).  The temperature fluctuaion is
reinforced due to nonlinear couplings between random set of magnetic
field fluctuations and an initial temperature perturbation, while
enhanced temperature fluctuations drive magnetic field fluctuations
linearly due to the baroclinic force. The nonlinear interactions
between MEDV magnetic fields keep mainitianing the resulting magnetic
structures.  Our numerical code employs a doubly periodic spectral
discretization of magnetic field and temperature fluctuations in terms
of its Fourier components, while nonlinear interactions are
deconvoluted back and forth in real and Fourier spectral spaces. The
time integration is performed by using the Runge-Kutta 4th order
method.  A fixed time integration step is used.  The conservation of
energy (8) is used to check the numerical accuracy and validity of our
numerical code during the nonlinear evolution of the magnetic field
and temperature fluctuations.  Varying spatial resolution (from
$128^2$ to $512^2$), time step ($10^{-2}, 5\times 10^{-3}, 10^{-3}$),
constant values of $K_n \lambda V_T^ 2/c^ 2 =0.1$ and
$(2/3)K_n \lambda - K_T \lambda =0.1$ are used to ensure the accuracy
and consistency of our nonlinear simulation results.  We also make
sure that the initial fluctuations are isotropic and do not influence
any anisotropy during the evolution. Anisotropy in the evolution can
however be expected from a $k_y=0$ mode that is excited as a result of
the nonlinear interactions between the finite frequency MEDV modes.

The magnetic and temperature fluctuations are initialized by using a uniform
isotropic random spectral distribution of Fourier modes concentrated
in a smaller band of wave number ($k<0.1~k_{max}$). While spectral
amplitude of the fluctuations is random for each Fourier coefficient;
it follows a $k^{-1}$ or $k^{-2}$ scaling.  Note again that our final results
do not depend on the choice of the initial spectral distribution. The
spectral distribution set up in this manner initializes random scale
turbulent fluctuations. Since there is no external driving mechanism
considered in our simulations, turbulence evolves freely under the
influence of self-consistent instability. Note however that driven
turbulence in the context of the MEDV mode will not change inertial range
spectrum to be described here.

\begin{figure}[t]
\begin{center}
\epsfig{file=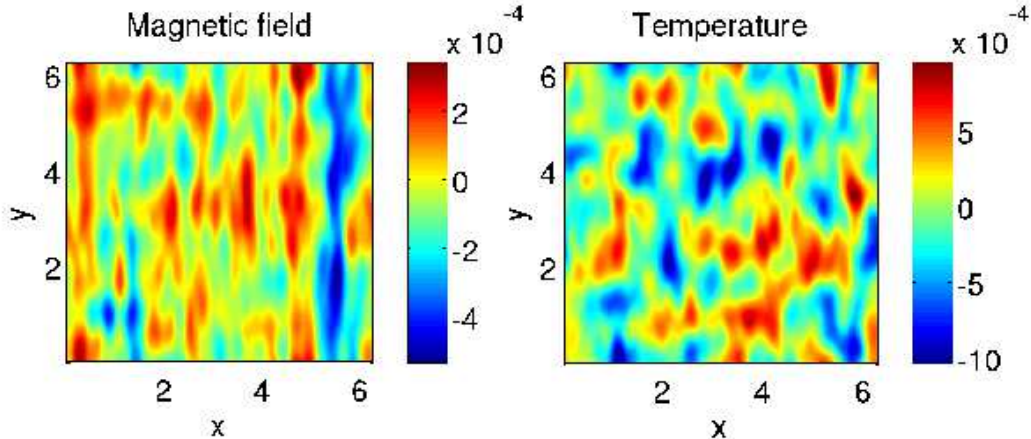, width=14.cm, height=6.cm}
\end{center}
\caption{Formation of streamer-like flows in nearly steady-state 
magnetic field fluctuations.  The temperature fluctuation is predominantly
nearly isotropic, but some eddies are aligned in the direction of the
background inhomogeniety. Numerical resolution is $512^2$, box
dimension is $2\pi \times 2\pi$, $\alpha = \beta = 1.0$.}
\end{figure}

 The nonlinear evolution of the magnetic and electron temperature
 fluctuations is governed typically by different nonlinearities in the
 two dynamical equations (5) and (6).  To understand the
 characteristics of the nonlinear interactions in our MEDV turbulence
 model, it is interesting to compare it with that of the
 Hasegawa-Mima-Wakatani (HMW) model [\cite{r13,r14,r15,r16}] describing
 the electrostatic drift waves in an inhomogeneous magnetoplasma.  For
 instance, the generalized magnetic field vorticity is driven by
 $\hat{z}\times \nabla B \cdot \nabla \nabla^2 B$ nonlinearity. The
 latter is similar to the ion polarization drift nonlinearity
 $\hat{z}\times \nabla \phi \cdot \nabla \nabla^2 \phi$, where $\phi$
 is the electrostatic potential, in a pseudo-3D HMW equation and
 signifies the Reynolds stress forces that play a critical role in the
 formation of zonal-flows [\cite{r8}]. Analogously, one can expect the
 nonlinear generation of electron flows in our MEDV model. The
 nonlinearity in the temperature perturbation equation (6) is
 $\hat{z}\times \nabla T_1 \cdot \nabla B$, which is akin to a
 diamagnetic nonlinear term in the HMW model. The role of this
 nonlinearity has traditionally been identified as a source of
 suppressing the intensity of the nonlinear flows in the drift wave
 turbulence. Nevertheless, the presence of the linear inhomogenous
 background in both the equations can modify the nonlinear mode
 coupling interactions in a subtle manner. Our objective here is to
 understand the latter in the context of the MEDV turbulent processes.
 The initial isotropic and homogeneous spectral distribution in
 magnetic field and temperature fluctuations, as described above,
 evolve dynamically following the set of equations (5) and (6). The
 spectral magnetic energy from large scale eddies migrates towards the
 smaller eddies following the Richardson's cascade law.  In
 configurational space, this essentially corresponds to breaking up
 the larger eddies into the smaller ones. Consequently, the mode
 coupling interactions in the Fourier space follow a Kolmogorov-type
 phenomenology in that spectral transfer, which predominantly occurs
 in the local $k$-space in the inertial range MEDV mode turbulence.
 During this process, each Fourier modes in the inertial range
 spectrum obey the vector triad constraints imposed by the vector
 relation ${\bf k} + {\bf p} = {\bf q}$. These nonlinear interactions
 involve the neighboring Fourier components (${\bf k},{\bf p},{\bf
 q}$) that are excited in the local inertial range turbulence.  The
 mode coupling interaction during the nonlinear stage of evolution
 leads to the formation of streamer-like structures in the magnetic
 field fluctuations associated with $k_y\approx 0, k_x \ne 0$. This is
 shown in Fig (1). Note that the streamer-like structures are similar
 to the zonal flows but contain a rapid $k_x$ variations, thus the
 corresponding frequency is relatively large.  The temperature
 fluctuations in Fig (1), on the other hand, depict an admixture of
 isotropically localized turbulent eddies and a few stretched along
 the direction of the background inhomogeneity.  We have performed a
 number of simulations to verify the consistency of our results in a
 strong turbulence regime.

 \begin{figure}[t]
\begin{center}
\epsfig{file=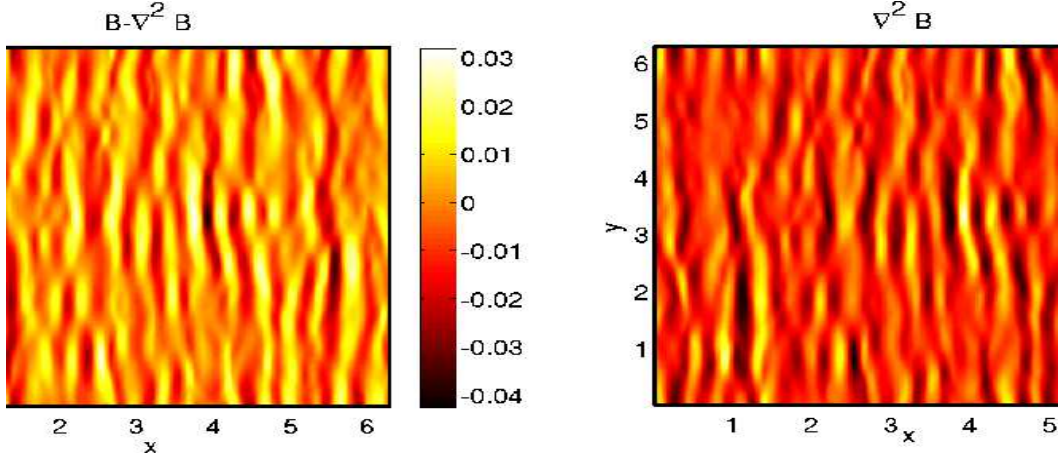, width=14.cm, height=6.cm}
\end{center}
\caption{Formation of relatively smaller scales streamer-like flows 
in generalized and magnetic field vorticity fluctuations corresponding
to the magnetic field structures shown in Fig (1).  As expected,
spectra of the latter contain smaller scales with increasingly dominated
$k_y \approx 0$ modes.}
\end{figure}

Figure (2) exhibits highly anisotropic nonlinear structures associated
with magnetic field vorticity ($\nabla^2 B$) and generalized magnetic
field vorticity ($B-\nabla^2 B$).  These structures, consistent with
Fig (1), vary along the $x$ direction and contain almost negligible
gradients along the direction of inhomogeneity.  The vorticities are
increasingly dominated by nonlinearly generated $k_y \approx 0$ modes.
The mode structures in Fig (2) comprises elongated (along the $y$
direction) smaller scale eddies because their spectra are propotional
to $k^2$ and $1+k^2$, respectively.

\section{Evolution of energy in the MEDV mode turbulence}

The formation of relatively smaller scale fluctuations in the steady
state MEDV mode turbulence (as displayed in Figs (1) and (2)) implies
essentially that the inertial range turbulent spectrum is dominated by
the higher Fourier modes. This is a characteristic of the forward
cascade processes in that the energy containing large scales eddies
transfer turbulent energy to the adjacent smaller scale until the
process of energy migration is terminated by dissipative processes. To
investigate the MEDV mode turbulent spectrum, we therefore include
dissipation (arising from the plasma resistivity and electron thermal
conduction involving electron-ion collisions) in the dynamical
equations (5) and (6) to damp smaller scales and thus terminate the
inertial range cascades.  Note however that nonlinear interactions, in
the absence of dissipation, conserves the total energy.

 \begin{figure}[t]
\begin{center}
\epsfig{file=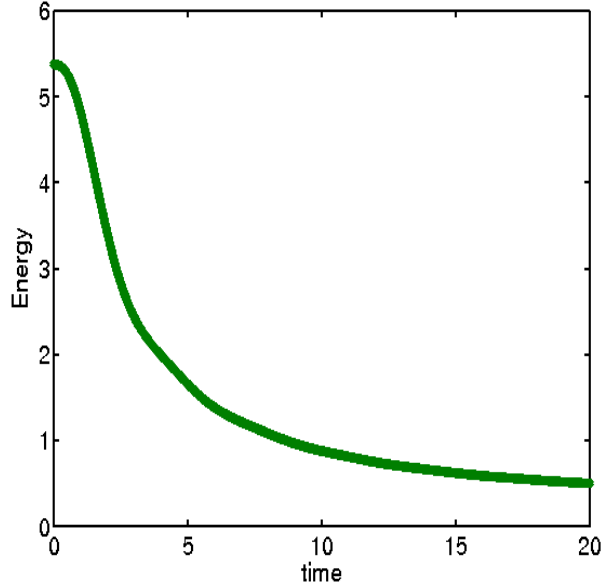, width=8.cm, height=8.cm}
\end{center}
\caption{Evolution of the total energy in the MEDV mode turbulence. A decay 
can be observed owing to the presence of small scale dissipation where much of the
turbulent energy is concentrated. The inclusion of dissipation not
only damps the smaller scales, but it also terminates the cascade
processes that results in an enhanced inertial range MEDV mode spectrum.}
\end{figure}

It is worth emphasising here that the two-component (i.e. in $B$ and
$T$) system of equations describing 2D MEDV mode turbulence is
characteristically different from the usual 2D drift wave or Euler
turbulence systems, especially in terms of the nonlinear evolution.
The presence of large scale structures have been routinely observed in
the latter. One of the reasons is the inverse cascade processes that
migrate spectral energy from the smaller to larger Fourier modes. By
contrast, in 2D MEDV mode turbulence the process of inverse cascade is
inhibited conspicuously by the diagmanetic-like nonlinearity that
depletes the large-scale flows. Furthermore, the interaction between
the nonlinearity and the background inhomogeneous flows (associated
$\partial_y T$ and $\partial_y B$ terms) leads to higly anisotropic
nonlinear streamer-like strucutres (see Figs (1) and (2)).
 
Since the energy is dominated by higher $k$'s in the MEDV mode
turbulence, the inclusion of dissipation is expected to damp the
turbulent energy.  This is shown in Fig (3), where a significant decay
of initial turbulent energy can be seen, which is followed by energy
saturation in the steady state. The decay of turbulent energy is
proportional to the dissipation. Thus larger is the dissipation,
higher the energy decay rates are.  After nonlinear interactions are
saturated, the energy in the turbulence does not dissipate further and
remains nearly unchanged throughout the simulations.  Correspondingly,
the energy transfer rate shows a significant change during the initial
and nonlinear phases.  However, when nonlinear interactions saturate,
the nonlinear transfer of the energy in the spectral space amongst
various turbulent modes becomes inefficient and the energy transfer
per unit time tends to become negligibly small, as shown in Fig (4).

 \begin{figure}[t]
\begin{center}
\epsfig{file=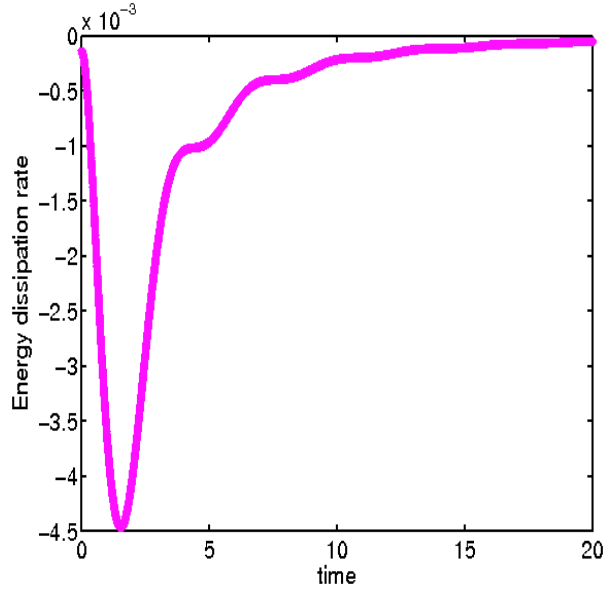, width=8.cm, height=8.cm}
\end{center}
\caption{Evolution of  energy dissipation rates in the MEDV mode turbulence.
During the inital decay, the dissipation rates vary rapidly followed
by a steady state where decay rates are nearly negligible and follows
a Kolmogorov-like energy cascade law. The latter is consistent with Fig (3)
that depicts a constancy of energy in the steady state MEDV mode turbulence. }
\end{figure}

\section{Energy spectrum}

It is clear from the nonlinear structures associated with $B$,
$\nabla^2B$ and $B-\nabla^2B$ that the mode structures are dominated
by $k_y\approx 0$ streamer-like modes.  Correspondingly, the inertial
range spectrum should contain signature of the highly anisotropic
streamer-like flows. One thus expects that the resultant energy
spectrum might deviate from the standard Kolmogorov spectrum that is
exhibited by the HM drift wave turbulence and its several variants. To
investigate the spectral features of MEDV mode turbulence, we plot the
inertial range energy spectrum in Fig (5). It is seen from this figure
that the MEDV mode spectrum is much steeper than its usual 2D drift
wave (isotropic) turbulence counterpart. The inertial range energy
spectrum, in the latter, exhibits a Kolmogorov-like $k^{-3}$ spectrum
[\cite{r17}] in the vorticity cascade regime (see the adjacent curve
in Fig (5)).  The spectral index associated with the MEDV mode
turbulence is close to $k^{-7}$ in the forward cascade vorticity
regime. There is a greater disparity in the two spectra. While the
disparity in the MEDV mode spectrum is caused essentially by the
anisotropic turbulent cascades, it is interesting to know whether it
can be understood from a Kolmogorov-like phenomenology. In view of
pursuing this issue, we derive the MEDV mode turbulence spectrum based
on the Kolmogorov-like dimensional arguments that are described in the
following.

 \begin{figure}[t]
\begin{center}
\epsfig{file=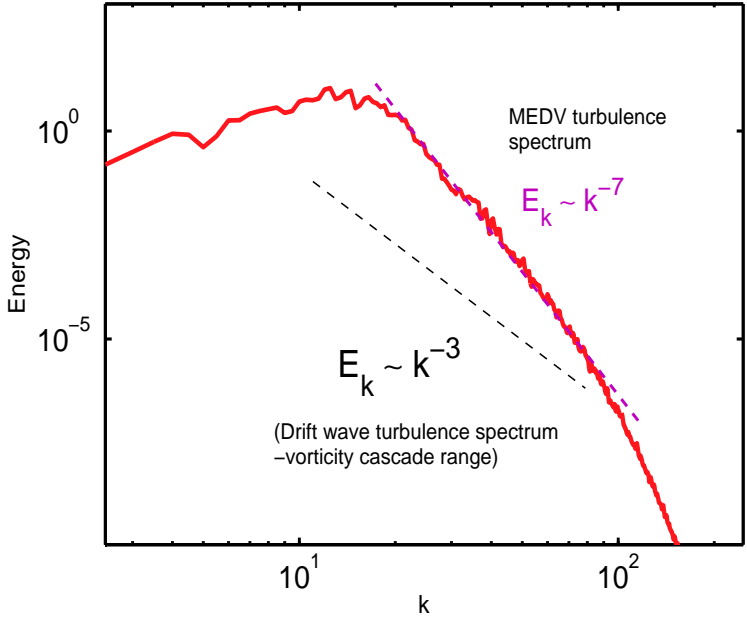, width=10.cm, height=9.cm}
\end{center}
\caption{MEDV mode turbulence spectrum is  much steeper than the usual
inhomogeneous drift wave turbulent spectrum in the forward cascade
vorticity regime ($k^{-3}$). The observed spectrum in our simulation
depicts a non Kolmogorov-like $k^{-7}$ spectrum. The underlying
steepness in the inertial range spectrum is mediated by the highly
anisotropic nonlinearly excited streamer-like structures for which
$k_y \approx 0$. }
\end{figure}

The energy cascade is mediated essentially by the nonlinear convective
forces in the MEDV mode turbulence and is proportional to $kB_k$,
where $k$ is a wavenumber associated with the magnetic field $B_k$ in
the spectral space. The corresponding energy transfer time can then be
estimated as $\tau_{nl} \sim (kB_k)^{-1}$. The energy transfer rates,
in the forward cascade vorticity regime, are determined by the
vorticity per unit nonlinear time, i.e.
\[ \varepsilon(k) \sim \frac{|\Omega_k|^2}{\tau_{nl}} \sim 
\frac{k^2 |B_k|^2}{\tau_{nl}}. \]
It should be borne in mind that the energy transfer rates, described
above, are influenced significantly by the presence of the propagating 
MEDV modes. The linear normalized frequency of this mode can be determined from the dispersion
relation. The linear interaction time of EDM with nonlinear eddies, in
$k \gg 1$ dispersive limit, can be given as $\tau_l \sim k/k_y$. This
interaction time needs to be necessarily included in the energy
transfer rates because the MEDV mode couples with nonlinear fluctuations. The
modified energy cascade rates are given in the following.
\[ \varepsilon(k)  \sim \frac{k^2 |B_k|^2}{\tau_{nl}^2} \tau_l \sim 
\frac{k^5|B_k|^3}{k_y}. \]
We next apply the Kolmogorov theory of turbulence that describes that
the inertial range spectral cascade is local and depends on the
wavenumber, as follows.
\[E_k \sim \varepsilon(k)^\alpha k^\beta.\]
On substituting the energy transfer rates in the above expression, one
obtains
\[k^{-1} |B_k|^2 \sim \left(\frac{k^5|B_k|^3}{k_y}\right)^\alpha k^\beta. \]
On equating the indices of identical bases, we get $\alpha=2/3$ and
$\beta=-20/3$. This further yields a $E_k \sim k^{-20/3}$ energy
spectrum in the forward cascade vorticity regime. The spectrum
obtained by including the MEDV mode interaction time in the Kolmogorov
phenomenology is in close agreement with (if not exactly identical to)
the observed energy spectrum (see Fig (5)). More steeper spectral
index can be followed {\it qualitatively} from the entire energy
spectrum that contains the anisotropic streamer-like flows. It can be
described by the following expression.
$$
E_k \sim \varepsilon(k)^{2/3}k^{-20/3} k_y^{-2/3}.
\eqno (9)
$$
Thus, the deviation from the usual Kolmogorov-like $k^{-3}$ spectrum
increases with the anisotropy.  The latter can be triggered either by
the presence of the mean magnetic field or background inhomogeneous
flows.  The spectrum described by Eq. (9) depicts a much steeper
spectral slope for streamer flows that have smaller magnitude of $k_y$
(or close to, but not entirely, zero).  Note carefully
that Eq. (9) does not reduce to a standard $k^{-3}$ spectrum
describing the isotropic MEDV fluctuations, because the anisotropy of
the spectral cascade is introduced in the energy cascade rates through
a ratio of linear and nonlinear periods. This choice does not allow us
to simply reduce the former to an isotropic turbulent spectrum. In this
respect, the anisotropic turbulent spectrum described by Eq. (9) is
only qualitative. A quantitative spectrum discerning the dependence on
$k_x$ and $k_y$ may require an altogether different analytic approach
and is beyond the scope of the present paper.

\section{Electron transport in magnetic structures}

We finally estimate the electron transport in the presence of magnetic 
field structures created due to nonlinear interactions between the 
MEDV modes. In the presence of an ensemble of magnetic field structures
the electrons perform a random walk and diffuse. Accordingly, 
there results and effective electron diffusion coefficient, $D_{eff}$,
which can be calculated from the cross correlation of the drift speed 
of the electrons in the turbulent magnetic field. We have  
\[ D_{eff} =
\int_0^\infty \langle {\bf V}_\perp({\bf r},t) \cdot {\bf
  V}_\perp({\bf r},t+ t^\prime) \rangle dt^\prime, \] 
where the angular bracket denotes ensemble spatial averages. The perpendicular
component of the electron fluid velocity associated with the MEDV mode
is ${\bf V}_\perp =\hat {\bf z} \times \nabla B$. Since the MEDV mode
turbulence is confined in the 2D plane, the effective diffusion coefficient
essentially describes the diffusion processes associated with the motion 
of electrons that produce fluctuating magnetic fields in the $x-y$ plane. 
We compute the evolution of the effective diffusion caused by nonthermal
magnetic fields of the MEDV modes.  This is shown in Fig (6). The effective 
diffusion is larger during the early phase of the simulations. This is where 
the isotropic MEDV mode turbulence progressively evolve towards anisotropy and
dominated increasingly by $k_y \approx 0$ modes. Once the anisotropic
streamers are formed, they begin to saturate such that no more energy
transfer takes place. The corresponding steady state transport also
shows saturation. This scenario is further consistent with the
evolution of energy and the subsequent decay rates depicted in Figs
(3) and (4), respectively.

 \begin{figure}
\begin{center}
\epsfig{file=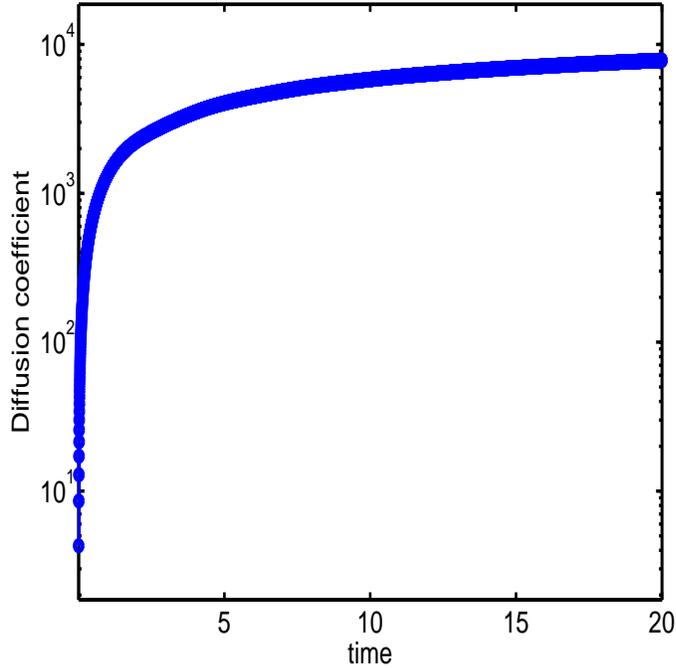, width=9.cm, height=9.cm}
\end{center}
\caption{The MEDV mode turbulence transport is dominated by large scale 
streamer-like flows that are dominated by $k_y \approx 0$ modes. The electron 
transport is higher initially and saturates in the steady state. }
\end{figure}

\section{Discussion}
In this paper, we have used computer simulations to investigate the
properties of 2D MEDV modes in a nonuniform plasma containing the
equilibrium density and electron temperature inhomogeneities. The
equilibrium is maintained due to a balance between the electric force
$-n eE_{0x}$ and the electron pressure gradient $\partial p_0/\partial
x$, where $p_0 =n (x) T_0(x)$. The electron temperature gradient can
create a phase shift between the magnetic field and electron
temperature perturbations, so that the MEDV mode grows provided that
$\eta =d{\rm ln}T_0/d {\rm ln} n > 2/3$. Large amplitude MEDV modes
interact among themselves and evolve in a turbulent state. The latter
consists of magnetic filaments/streamers and localized electro heated
regions in nonuniform plasmas. The resulting turbulent spectrum
deviate from the usual Kolmogorov [\cite{r17}] or the Fyfe-Montgomery
[\cite{r18}] scalings, which describe the complex fluid turbulence and
drift wave turbulence in plasmas, respectively.  Due to the random
walk process, electrons would diffuse in the magnetic field structures
that are created due to the nonlinear interactions between the MEDV
modes.  The magnetic field structures may also cause the electron heat
transport on account of the electron temperature perturbation which
are driven by the localized magnetic fields of the MEDV modes. The
present investigation should help to understand the generation of
flows, structures and associated spectra producing anomalous transport
of particles and heat in inertial confinement fusion plasmas.

\label{lastpage}

\begin{thereferences}{9}
\bibitem{r1}
 Jones, R. D. 1983 Magnetic Surface Waves in Plasmas.
{\it Phys. Rev. Lett.} {\bf 51}, 1269.

\bibitem{r2}
 Yu, M. Y. and Stenflo, L. 1985 
Magnetic surface wave instabilities in plasmas.
{\it Phys. Fluids} {\bf 28}, 3447.

\bibitem{r3}
Stenflo, L.,  and  Yu, M. Y. 1986
Instabilities of baroclinically driven magnetic and acoustic waves.
 {\it Phys. Fluids} {\bf 29}, 2335.

\bibitem{r4}
Stenflo, L., Shukla, P. K., and Yu, M. Y. 1987
Decay of magnetic-electron-drift vortex modes in plasmas.
{\it Phys. Rev. A} {\bf 36}, 955.

\bibitem{r5}
Shukla, P. K., Yu, M. Y., and Stenflo, L. 1988 
Stimulated Compton scattering of magnetic-electron-drift vortex waves off plasma ions.
{\it Phys. Rev. A} {\bf 37}, 2701.

\bibitem{r6}
Yu, M. Y., Shukla, P. K., and Spatschek, K. H. 1974
Scattering and modulational instabilities in magnetized plasmas.
 {\it Zh. Naturforsch. A} {\bf 29}, 1736;
Shukla, P. K., Yu M. Y., and Spatschek, K. H. 1975
Brillouin backscattering instability in magnetized plasmas.
 {\it  Phys. Fluids} {\bf 18}, 265;
Shukla, P. K. 1978 
Modulational instability of whistler-mode signals.
{\it Nature} {\bf 274}, 874 (1978);
Stenflo, L. 1970 Kinetic theory of three-wave interaction in a magnetized 
plasma. {\it J. Plasma Phys.} {\bf 4}, 585.

\bibitem{r7}
Sharma, R. P.,  and  Shukla, P. K. 1983
Nonlinear effects at the upper-hybrid layer.
{\it  Phys. Fluids} {\bf 26}, 87;
Murtaza, G. and  Shukla, P. K. 1984
Nonlinear generation of electromagnetic waves in a magnetoplasma.
{\it J. Plasma Phys.} {\bf 31}, 423;
Shukla, P. K., and Stenflo, L. 1985
Nonlinear propagation of electromagnetic waves in magnetized plasmas.
 {\it Phys. Rev. A} {\bf 30}, 2110.

Stenflo, L., and Shukla, P. K., 2000
Theory of stimulated scattering of large-amplitude waves.
{\it J. Plasma Phys.} {\bf 64}, 353.

\bibitem{r8}
Shukla, P. K., Yu, M. Y., Rahman, H. U., and  Spatschek, K. H. 1981
Excitation of convective cells by drift waves.
{\it Phys. Rev. A} {\bf 23}, 321;  
Nonlinear convective motion in plasmas.
{\it Phys. Rep.} 1984
{\bf 105}, 227;
Shukla, P. K.,  and Stenflo, L. 2002
Nonlinear interactions between drift waves and zonal flows.
{\it Eur. Phys. J. D} {\bf 20}, 103. 

\bibitem{r9a}
Nycander, J., Pavlenko, V. P., and Stenflo, L. et al, 1987 
Magnetic vortices in nonuniform plasmas. 
{\it Phys. Fluids} {\bf 30}, 1367. 

\bibitem{r9}
Shukla, P. K.,  Birk, G. T., and  Bingham, R. 1995
 Vortex streets driven by sheared flow and applications to black aurora.
{\it Geophys. Res. Lett.} {\bf 22}, 671.

\bibitem{r10}

Stenflo, L. 1987 
Acoustic solitary vortices. 
{\it Phys. Fluids} {\bf 30}, 3297.

\bibitem{r11}
Jucker, M., and  Pavlenko, V. P. 2007
On the modulational stability of magnetic structures in electron drift turbulence.
 {\it Phys. Plasmas} {\bf 14}, 102313.

\bibitem{r12}
Andrushchenko, Z. N.,  Jucker, M., and  Pavlenko, V. P., 2008,
Self-consistent model of electron drift mode turbulence.
{\it J. Plasma Phys.} {\bf 74}, 21.

\bibitem{r13}
Hasegawa, A.,  and  Mima, K. 1977
Stationary spectrum of strong turbulence in magnetized nonuniform plasma.
{\it Phys. Rev. Lett.} {\bf 39}, 205;
Mima, K.,  and  Hasegawa, A. 1978 
Nonlinear instability of electromagnetic drift waves.
{\it Phys. Fluids} {\bf 21}, 81.
\bibitem{r14}
Hasegawa, A.,  and Wakatani, M. 1983  
Plasma Edge Turbulence.
{\it Phys. Rev. Lett.} {\bf 50}, 682; 
{\it ibid.} {\bf 59}, 1581 (1987). 
\bibitem{r15}
 Hasegawa, A. 1985
Self-organization processes in continuous media.
 {\it Adv. Phys.} {\bf 34}, 1.
\bibitem{r16}
Horton, W.,  and Hasegawa, A. 1994 
Quasi-two-dimensional dynamics of plasmas and fluids.
{\it Chaos} {\bf 4}, 227.
\bibitem{r17}
 Kolmogorov, A. N. 1951 
The Local Structure of Turbulence in Incompressible Viscous Fluid for Very Large Reynolds' Numbers.
{\it C. R. Acad. Sci. U. R. S. S.} {\bf 30}, 301,
and {\bf 30}, 538 (1941).
\bibitem{r18}
Fyfe, D.,  and  Montgomery, D. 1979 
Possible inverse cascade behavior for drift-wave turbulence.
{\it Phys. Fluids} {\bf 22}, 246.
\end{thereferences}
\end{document}